\documentclass[twocolumn,noshowpacs]{revtex4}

\usepackage{graphicx}%
\usepackage{dcolumn}
\usepackage{amsmath}

\makeatletter
\def\btt#1{\texttt{\@backslashchar#1}}%
\DeclareRobustCommand\bblash{\btt{\@backslashchar}}%
\makeatother

\usepackage{hyperref}

\begin{document}


\title{Measurement of attenuation rather than phase: discrepancy between Schr\"odinger vs non-relativistic Klein-Gordon equations}


\author{Frank V. Kowalski}
\affiliation{Physics Department, Colorado
School of Mines, Golden CO. 80401 U.S.A.}

\begin{abstract}
In classical mechanics adding a constant to the energy has no physical effect. In quantum mechanics, this alters the phase velocity of the wave. The two theories that describe non-relativistic quantum mechanics should agree on their predictions but they have vastly different phase velocities; one (non-relativistic Klein-Gordon eq.) adds $mc^{2}$ to the energy, while the other (Schr\"odinger eq.), does not. Unlike the former, only the latter allows an attenuator to outpace free particle wave crests. Nevertheless, an attenuator can outpace a wave group in both cases. Yet, only the latter exhibits multiple attenuations for such an attenuator trajectory. These differing predictions cast doubt on the validity of either adding a rest energy term or not, illustrating a relativistic effect within ``non-relativistic'' quantum mechanics.
\end{abstract}


\maketitle

\section{Introduction}

The Schr\"odinger eq. is a non-relativistic approximation of the Klein-Gordon eq. for spin-0 particles.\cite{feshback} Similarly, extending the Schr\"odinger eq. to include spin yields the non-relativistic approximation of the Dirac eq.\cite{tsutsui}  At high energies, however, the single-particle interpretation of relativistic wave eqs. becomes invalid. They struggle to handle particle creation and annihilation, result in the Klein paradox, and fail to conserve total probability.\cite{tsutsui}

The Schr\"odinger eq. has been enormously successful in understanding the structure of matter while the Klein-Gordon eq. is foundational in quantum electrodynamics. Despite these successes, empirical gaps persist in their relationship. One such case involves the non-trivial motion of an attenuator along the trajectory of the wavefunction.

Measurement of either attenuation or shifts in phase indicate quantum rather than classical behavior. However, quantum applications primarily utilize phase shifts measured with an interferometer due to their acute sensitivity to external variables. Both decoherence and path dependent attenuation reduce the visibility of this interference. However, unlike most phase shift measurements, the effects of attenuation can be determined outside of an interferometer.

One difference between the Schr\"odinger and non-relativistic Klein-Gordon eqs. is their phase velocities, which for a free particle in an eigenstate of momentum in the Schr\"odinger eq. is $\omega/k=KE/p=v_{g}/2$ where $KE=p^{2}/2m, p$, $m$, and $v_{g}$ are the kinetic energy, momentum, mass, and group or particle velocity, respectively. In the non-relativistic limit the phase velocity in the Klein-Gordon eq. is, in this case, $\omega/k=(mc^{2}+KE)/p \approx c^{2}/v_{g}$ where $c$ is the speed of light.

This difference in phase velocities arises from the rest energy term present in the Klein-Gordon eq. but not in the  Schr\"odinger eq. Using the ansatz $\Phi=\exp[-imc^{2}t/\hbar] \Psi$ in the Klein-Gordon eq. and the assumption that the free particle kinetic energy is much less than the rest energy results in $\Psi$ satisfying the Schr\"odinger eq.\cite{feshback} Nevertheless, the frequency of the free particle wavefunction in the non-relativistic Klein-Gordon eq. is $(mc^{2}+m v_{g}^{2}/2 )/\hbar$.

The group velocity, $d\omega/dk$, for the Schr\"odinger eq. is the same as that for the Klein-Gordon eq. since their frequencies differ only by a constant rest energy term. Since each spinor component satisfies the Klein-Gordon eq., these phase and group velocities also apply to the Dirac eq..

All of the Fourier components of a Gaussian wave group in the Klein-Gordon formalism therefore contain the global phase factor, $\exp[-i mc^{2}t/\hbar$]. Since the probability density function (PDF) is defined as $|\Psi|^{2}$, this factor vanishes in the PDF, yielding the Schr\"odinger PDF. A positive-definite probability is therefore conserved in this non-relativistic approximation to the Klein-Gordon eq. Although both wave groups have the same PDF shape, wave crests move through them at vastly different speeds.

The dissimilarity between phase and group velocity was first discussed by Lord Rayleigh.\cite{rayleigh}. Following the theory of relativity interest was focused on the limits of information transfer in wave phenomenon. Initially, this involved electromagnetic waves propagating through a medium in its anomalous dispersion region.\cite{sommerfeld,brillouin} A similar treatment for the Klein-Gordon eq. occurred later.\cite{schrader} Current interest in the speed of a wave group involves evanescent waves in quantum tunneling \cite{sharoglazova} and in generating extremely slow electromagnetic pulses in a medium.\cite{stiller} Interest in the connection between the Schr\"odinger and Klein-Gordon eqs. continues.\cite{gourdain} Resolution of the negative PDF for the Klein-Gordon eq. has also been discussed.\cite{lin}

The area of focus here, however, involves the dynamics of a thin broad-band attenuator moving along the direction of propagation of Schr\"odinger and non-relativistic Klein-Gordon wavefunctions. There is a distinct lack of prior research addressing this topic. Peripheral discussions focused on the interaction of electromagnetic waves with a moving beamsplitter (BS), which naturally attenuates its outputs. This includes both longitudinal (parallel to the wavevector) \cite{xuereb} and transverse BS motion.\cite{stefanov} Accelerated motion of a pulsed BS for atom interferometry has been studied, although the emphasis is on the measurement of phase shifts rather than on that of wavefunction attenuation.\cite{debavelaere}

Coherence-preserving BSs have been successfully applied in reducing the amplitude of matter waves, including those of neutrons \cite{rauch}, heavy molecules, macromolecules, and Bose-Einstein condensates.\cite{cronin}  They have been shown to remain coherent despite acceleration. \cite{debavelaere,frank} The most common approach relies on matter-wave diffraction produced by a mechanical or optical grating. The zeroth order beam from such a grating is attenuated by loss of probability flux into higher diffracted orders. However, the energy of individual particles in the zeroth-order beam remains constant, regardless of motion of the BS.

A grating can also be formed by physically removing particles from the beam in a transverse periodic pattern. \cite{arndt} Such particle removal or absorption, when executed uniformly over the full transverse beam profile, serves as another example of the attenuator described below.

Other attenuator/BS methods rely on mechanisms such as a plate (modeled as rectangular potential barriers), dissipative or non-dissipative nanostructures, polarization, electric or magnetic fields, reflection, or Bragg scattering.\cite{cronin} These act on the center of mass of the particle.  Alternatively, some methods to split the wavefunction modify the particles internal states. Examples are a spin dependent optical BS, \cite{spin}, a dark and bright state BS. \cite{bright}, a Rabi oscillation BS \cite{rabi1,rabi2}, and a Stern-Gerlach BS. \cite{stern}

The discussion below centers on a macroscopic BS/attenuator. However, as its mass decreases, many-body quantum effects become more detectable. Attenuation generated from a particle traversing a rectangular potential barrier is an example. Reflections at the two boundaries of the barrier generate two-body quantum correlated interference. \cite{kowalskiedu}  Reduced probability of particle transmission correlates with enhanced probability of barrier recoil. Both attenuator systems—the particle-barrier and particle-grating— exist in a superposition of two states: a transmitted particle with a non-recoiled BS, and a reflected particle with a recoiled BS.

Consider an overview of the proposed methodology. Let a particle in an eigenstate of momentum traverse a thin grating BS when it is initially at rest. We follow a single wave crest that starts out attenuated in this way. Next, the BS moves along the wave crest's path of motion, passing through it before coming to rest. This process weakens the wave crest in three distinct stages: at the initial and final BS rest positions and as the BS overtakes the wave crest. Only the initial attenuation occurs for a Klein-Gordon wave crest since the BS can never overtake this wave crest. For the same reason, such motion-induced attenuation is also absent for electromagnetic waves.

The BS leaves the phase of the wave unaffected, yielding identically shaped PDFs for a wave group in both formalisms. Therefore, a wave group will initially traverse the BS, then the BS will overtake the wave group, and finally the BS will come to rest in front of the wave group in both formalisms (even though a wave crest cannot be overtaken in the Klein-Gordon formalism). The difference between the Schr\"odinger and Klein-Gordon eqs. is not a phase shift or location of the wave group during this BS trajectory but is only the attenuation of the wave group.

Wavefunction attenuation for this BS trajectory is a uniquely quantum effect. Unlike in a typical phase measurement, measurement of attenuation does not require both a BS and an apparatus whose mechanical stability is comparable to the extremely short coherence lengths associated with massive particles. Owing to the phase-insensitivity of the attenuation measurement, sequential particles passing through the BS undergo identical attenuation independent of their relative phases or coherence lengths.

Surpassing the Schr\"odinger phase velocity of non-zero rest mass particles with an attenuator is not unreasonable. Neutrons can be generated at speeds of $10 \rightarrow 1000$ m/s for cold/thermal beams. \cite{ahmed} With laser cooling the speeds of atoms can be reduced to less than $30$ m/s. \cite{hosoya} BSs have been used to generated interference in these systems.

The initial discussion below involves motion of an attenuator within a Sagnac interferometer. Such interferometers utilizing non-zero rest mass particles have been constructed for electrons\cite{zivkovic}, neutrons\cite{kawasaki}, and atoms.\cite{zhou}

When the coherence length of a wave packet exceeds the path difference in an interferometer, its interference pattern depends predominately on the central momentum of the wave group. In this regime, the interference generated by the wave group effectively behaves as would a momentum eigenstate for path variations that are much smaller than the coherence length. The following analysis first examines this specific case, expands to wave groups, and ultimately addresses the effect in the absence of an interferometer.

The significance of the phase velocity in interferometry is often concealed by using a snapshot method for determining the difference in phase; time is fixed while the number of wave crests is compared along the two paths. This method is often used for determining phase shifts in atom interferometry. \cite{cronin,zhang} Methods of calculating the phase when external forces act on the atom include the Feynman path integral \cite{storey} and the atom optics ABCD formalism. \cite{borde}

The retarded phase method, however, emphasizes the importance of the phase velocity in such calculations. \cite{kowalski2010} For example, let the phase at the input port of a static interferometer be given by $-\omega_{0} t$. The phase at the output port along one path is then $-\omega_{0} t_{ret}$, where $t_{ret}$ is the retarded time for that path (the time at which the wave crest at the output port was at the input port). For an electromagnetic wave traversing this interferometer $t_{ret}=t-t_{out}$, where the wave crest transit time to the output port is $t_{out}=L/c$ and $L$ is the distance that it travelled. In the Klein-Gordon formalism these transit times are extremely small yet the phase, a product of the the extremely large frequency with this time, is the same as that for the Schr\"odinger eq.

One advantage of the retarded phase method is that the phase along a path whose length depends on time is easily determined. An example involves reflection of a particle in an eigenstate of momentum from an accelerating mirror. The reflected states for the Schr\"odinger and the Klein-Gordon eqs. then differ in the non-relativistic limit due to the difference in  phase velocities (although they don't differ for constant mirror velocities).\cite{kowalski2010}

\section{Results}

\subsection{Momentum eigenstates}

\begin{center}
\begin{figure}
\includegraphics[scale=0.41]{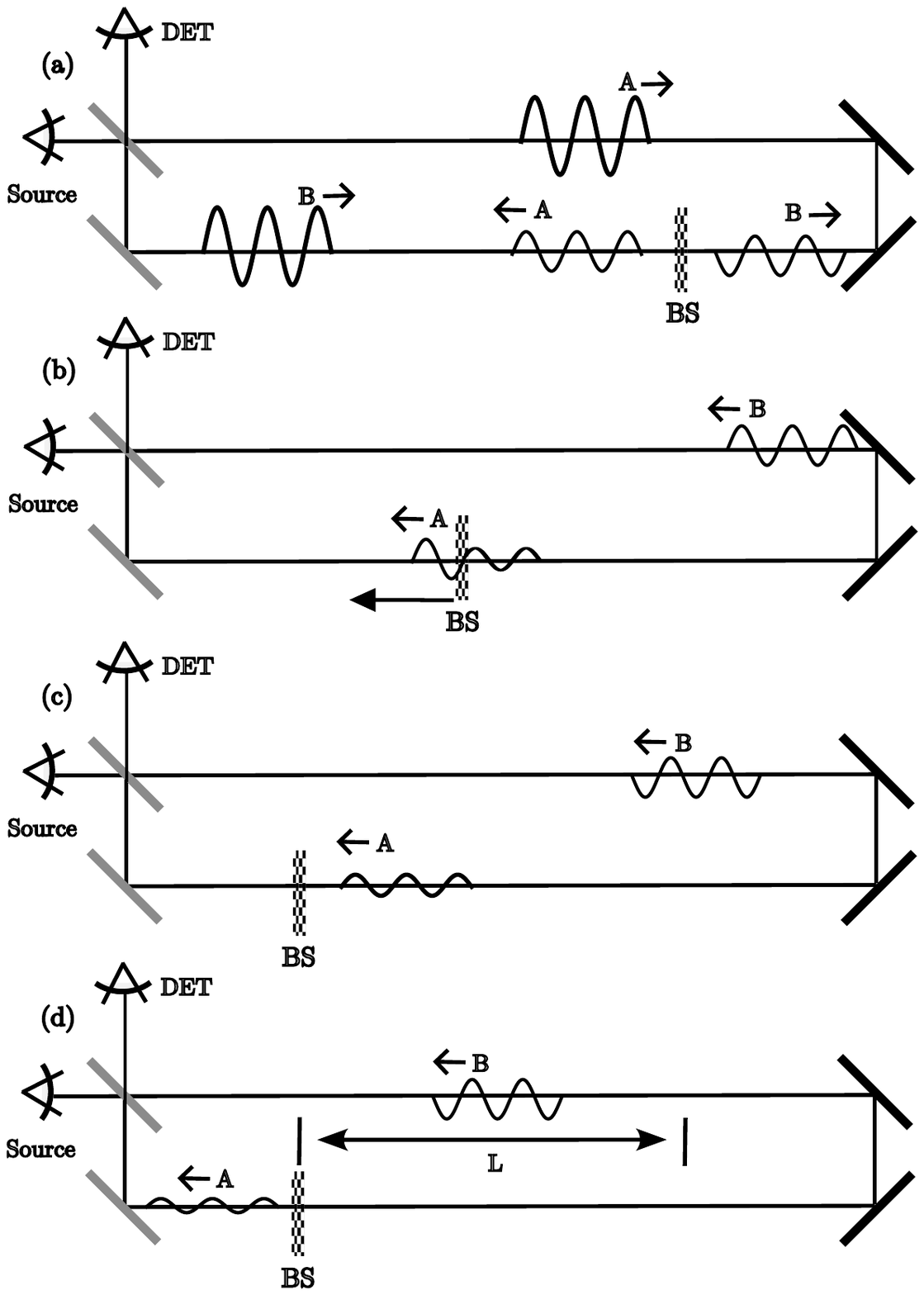}
\caption{Interference generated by one dimensional motion of an attenuator, BS, within a Sagnac interferometer for the  Schr\"odinger eq.. Only segments of the wave trains are shown. The highest amplitude segments in (a) illustrate the counter-propagating waves prior to traversing the attenuator. The length of the arrows indicate the speeds of the wave crests and attenuator while the amplitude of a wave segment indicates how often it traversed the attenuator. The source, S, emits a harmonic wave that interacts with a stationary attenuator in frame (a). The attenuator then moves to the left faster than the CW traveling wave in frame (b) generating a transmitted wave that is left behind it. The attenuator then comes to rest after having passed through the initially transmitted CW sequence of wave crests (shown in frame(a)) in frame (c). This lagging wave crest sequence then traverses the attenuator for a final time after it comes to rest, reducing its amplitude again. The attenuator travels a distance $L$.}
\label{fig1}
\end{figure}
\end{center}

The retarded phase method can reveal interferometry that might be non-intuitive when considering only the snapshot method. An example for the  Schr\"odinger eq. is the one-dimensional nontrivial motion of a attenuator that exceeds the phase velocity on part of its trajectory, as shown in the lab frame schematically in fig. \ref{fig1}. The linear motion of this attenuator, indicated as BS in this fig., is constrained to be within the Sagnac interferometer. The attenuator is initially treated as an ideal (zero-thickness, achromatic) element, with more realistic models presented below. It is initially stationary as shown in fig. \ref{fig1}(a) then moves faster than the speed of a wave crest in fig. \ref{fig1}(b) before coming to rest in fig. \ref{fig1}(c).

The eigenstate of momentum injected into this interferometer is represented only by wave train segments for clarity. For example, the two largest amplitude wave crest segments shown in fig. \ref{fig1}(a), labeled $A$ and $B$, represent the clockwise (CW) and counterclockwise (CCW) wavefunctions before traversing the attenuator. The segments labeled $A$ and $B$ with reduced amplitudes represent the waves after traversing the attenuator. The grating attenuator directs diffracted beams away from the detector. The effect of reflections (not shown in the diagram) from a plate attenuator are suppressed by anti-reflection coatings or by tilting the attenuator to misalign back-reflections away from the detector.

The CW wave segment overtaken by the attenuator experiences three traversals of the attenuator, compared to a single traversal for the CCW segment. Since the difference in path lengths between the CW and CCW paths is zero the particle beam intensity at the output of the Sagnac interferometer, after the attenuator has come to rest, is given by $I_{Sch}=I_{ccw}+I_{cw}+2 T^{2} \sqrt{I_{ccw}}\sqrt{I_{cw}}$, where $T$ is the attenuator transmission coefficient and $I_{ccw}$ and $I_{ccw}$ are the intensities for one traversal of the attenuator. The interference of these counter propagating waves with different attenuations lasts only until all the wave crests that were left behind the moving attenuator have transmitted through it again when it comes to rest. The output then returns to that for no motion of the attenuator $I_{Sch}^{0}=I_{ccw}+I_{cw}+2 \sqrt{I_{ccw}}\sqrt{I_{cw}}$.

No such variation in the interference due to motion of the attenuator can occur for wave crests of electromagnetic radiation or for those obeying the Klein-Gordon eq. since the speed of the attenuator can never exceed that of a wave crest. The CW and CCW waves then traverse the attenuator only once for these wave eqs.

Other examples where such an attenuator could move through waves involve ultra slow electromagnetic pulses or acoustics. However, both rely on wave propagation through material media, which is inherently disrupted by the movement of an attenuator.

The quantum behavior of the attenuator, described above, can be derived using reflection and transmission from a thin barrier potential (for a thick barrier potential the phase transit time depends on the width of the barrier as discussed later). Two approaches are used: (1) transformation to and from the rest frame of the BS, or (2) the application of the retarded phase method in the lab frame. The former method is straightforward to analyze since it only involves static boundary conditions. However, this transformation to and from the attenuator frame for the  Schr\"odinger eq. requires a unusual phase factor.\cite{levi}

During most of its trajectory the attenuator is either at rest or moving at constant velocity. A small duration, however, involves acceleration, the effects of which are approximated as a sequence of momentarily co-moving inertial frame interactions between the wave and the attenuator. The calculation utilizing these transformations between the lab and attenuator frame is outlined next.

Consider first the interaction of the wave with the attenuator for the Schr\"odinger eq. while the attenuator moves at constant speed $V$ (after the wave has traversed the stationary attenuator). The attenuator-wavefunction interaction occurs in two ways in fig. \ref{fig1}: (I) when both the attenuator and incident wave crests move in the same CW direction and (II) when the moving attenuator and the CCW wave crests move in opposite directions.

In case (I) the CW wave crests that move in the same direction as the attenuator in the lab frame, as shown in fig. \ref{fig1} (b), now move toward the attenuator in its frame. Continuity of the incident, transmitted, and reflected wave functions and their first derivatives with the barrier is trivially satisfied in the attenuator/barrier frame. The results are then transformed back to the lab frame where the attenuated transmitted wave crests now move slower than and behind the attenuator in the direction of attenuator motion. These lagging wave crests have the same frequency, $mv_{g}^{2}/(2 \hbar)$, and wavevector, $mv_{g}/ \hbar$, as those of the wave before interacting with the BS.

After the attenuator comes to rest the wave crests, that were attenuated twice, transmit through the attenuator yet again for a total of three times before reaching the detector. as shown in \ref{fig1} (d). The reflected wave crests are not considered since they never reach the detector.

Consider case (II) in the frame of the BS. The incident wave crests (ccw) move faster toward the attenuator than in case (I). The reflected crests are again disregarded since they do not enter the detector due to the orientation of the BS. The wave crests transmitted through the attenuator in the lab frame have the same frequency, $mv_{g}^{2}/(2 \hbar)$, and wavevector, $mv_{g}/ \hbar$, as before interacting with the attenuator while moving in the direction opposite to that of the BS. Their amplitude is that associated with one transmission through the BS.

Contrast this calculation with that using the Klein-Gordon eq.. Boundary conditions are applied in the attenuator/barrier frame after a Lorentz transformation from the lab frame. After interaction, the wave's parameters are transformed back to the laboratory frame, where the reflected phase velocity is $c^{2}/(2V+v_{g})$. This result (in addition to that for the same interaction in the Schr\"odinger formalism) can also be obtained using the retarded phase method without the need for a transformation to and from the frame of the BS.\cite{kowalski2010} However, it is only the transmitted wave that is of concern here.

Consider the moving attenuator in the lab frame for case (I) in the Klein-Gordon eq.. The wave crests that transmit through the attenuator while it is stationary cannot again transmit through the moving attenuator since the attenuator cannot overtake them. These wave crests therefore remain unaffected in both amplitude and phase by the motion of the BS. Similarly, in case (II) the wave crests transmit through the attenuator only once. The particle beam intensity at the output of the Sagnac interferometer, $I_{KG}=I_{ccw}+I_{cw}+2 \sqrt{I_{ccw}}\sqrt{I_{cw}}$, the same as that for a stationary BS. This result also applies for electromagnetic waves.

\subsection{Plate and grating BSs }

A plate attenuator/BS provides an illustration of deviations from ideal (thin) BS behavior. Retro-reflection occurs at the interfaces, while transmission propagates through the plate's medium (stationary or moving), characterized by an index of refraction. A neutron propagating through a plate that moves at constant speed is an example. This is modeled as a moving rectangular barrier potential in the Schr\"odinger eq., resulting in the neutron Fresnel-Fizeau effect.\cite{horne}

In addition to the different attenuations for the transmitted counterpropagating waves, discussed in the previous section, the time for a wave crest to traverse such a moving plate BS is longer in the clockwise than in the counterclockwise direction resulting in different Fresnel-Fizeau phase shifts.

In the Klein-Gordon formalism a wave crest traverses the plate when it is at rest. A later wave crest in this wave train traverses the plate (since the plate can never overtake the earlier wave crest) when it is moving and a yet later wave crest of this wave train traverses the plate after it finally comes to rest. All of these sequential wave crests experience only one attenuation resulting in only one attenuation for the wave train. Yet, the transit times during plate motion still differs between the counterpropagating waves (the Fresnel-Fizeau effect), resulting in the same phase shift predicted for plate motion by the Schr\"odinger eq.

The effect on the transmitted wave from accelerating the plate to sub-phase velocity speeds has been calculated\cite{kowalski1993} and verified experimentally.\cite{frank} This results in a frequency shift of the transmitted wave crests, $\Delta \nu=m g L (1-n)/n \hbar$, where $m$, $g$, $L$, and $n$ are the neutron mass, plate acceleration, plate length, and plate index of refraction, respectively.

This system was studied to examine the validity of the Equivalence Principle, specifically by showing that wave propagation within an accelerated assembly (source, plate, and detector) is indistinguishable from that in a uniform gravitational field. A related experiment involves dropping the plate in a uniform gravitational field. The effect of acceleration is then related to a displacement of the plate in its inertial frame\cite{balzas} as it falls freely; the loss in potential energy due to this displacement in the gravitational setting results in an increase in frequency of the wave traversing the plate. In a similar setting, plate displacement measurements have been made of order the Planck length when interacting with electromagnetic radiation in a gravitational field.\cite{kowalski1993b,kowalski2022}

Acceleration of the plate in the direction (or in the opposite direction) of wave crest motion increases (or decreases) the frequency of the transmitted wave. The transmitted wave train then acquires frequency chirps during the acceleration and deceleration stages of the plate trajectory. The frequency of the transmitted wave train during the constant speed part of the trajectory does not change.\cite{horne} These finite segments of frequency chirps then disperse as they move through the wave train. The frequency of the wave crests (or speed of the transmitted particle during plate acceleration) increases or decreases for $n>1$ or $n<1$, respectively.

A thin grating BS avoids these accelerating matter complications since the phases of the incident and zeroth order diffracted beams are the same at the grating apertures. The zeroth order beam is attenuated via loss into higher order diffracted beams. In fig. \ref{fig1}(a) the zeroth order beam transmits through the grating generating the CW wavetrain that is later overtaken by the grating attenuator in fig. \ref{fig1}(b). After the grating comes to rest the wavetrain left behind it then transmits in zeroth order through the grating again (this wavetrain has then traversed the grating three times) and interferes with the zeroth order CCW beam (that has traversed the grating one time). The grating is oriented so that specularly reflected and diffracted light is directed away from the detector.

During the acceleration phase, the speed of the attenuator passes through that of the phase velocity in the the Schr\"odinger formalism. The wavelength in the frame of the attenuator increases. Higher-order diffracted beams are eliminated in a grating attenuator, due to this increase in wavelength, and more energy is redirected into the zeroth-order transmission. This induces a transient spike in grating transmission. Such an effect does not occur in the Klein-Gordon formalism since the speed of the attenuator can never exceed that of a wave crest. A thin-plate achromatic attenuator prevents this transient effect.

This system is modeled as a superposition of two wavefunctions in the Schr\"odinger eq.: the steady-state transmission through the attenuator and the transient increase in transmission due to the accelerating attenuator when it overtakes a wave crest. Because the transient component travels at the group velocity (faster than a wavecrest) this wavefunction component can be separated from the effect illustrated in fig. \ref{fig1}.

\subsection{Superposition of momentum eigenstates }

Consider next motion of a wave group through the Sagnac interferometer in the Schr\"odinger formalism. Let a free particle of mass $m$, that is in a Gaussian wave group state of width $\sigma \ll L$, be launched into the interferometer. $L$ is the distance between the initial and final positions of an achromatic attenuator during its trajectory. This attenuator begins its motion when the CW wave group has just finished traversing the attenuator and finishes its motion after the attenuator has overtaken the wave group.

The Fourier components of the CW wave group that have phase velocities greater than that of the attenuator will only experience one attenuation from the BS. However, those with phase velocities less than the velocity of the attenuator will experience attenuation three times resulting in a distorted wave group at the interferometer output port. When the attenuator moves faster than the phase velocities of most Fourier components, this effect becomes negligible, allowing the transmitted wave group to pass through attenuated but undistorted.

The wave group interference at the output port for the Schr\"odinger eq. shifts from $I_{Sch}^{0}$ when the attenuator remains stationary to $I_{Sch}$ when it follows the trajectory described above. For the Klein-Gordon eq. the wavecrests of all the Fourier components experience only one transit through the BS. The wave group interference is given by $I_{KG}$ and is independent of the attenuator trajectory.

\begin{center}
\begin{figure}
\includegraphics[scale=0.41]{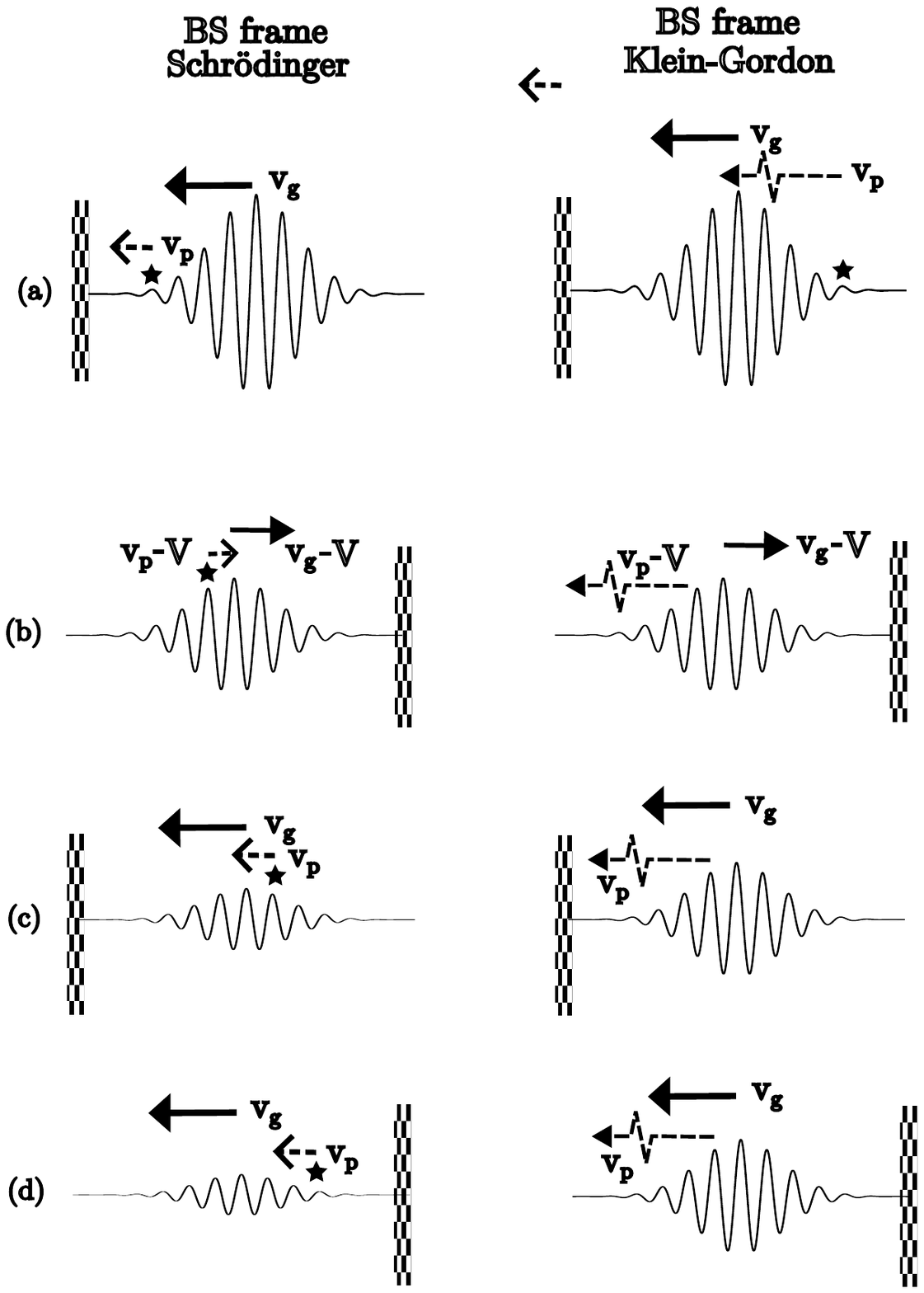}
\caption{Interaction of a wave group along the attenuator trajectory of fig. \ref{fig1}, as viewed in the rest frame of the attenuator/BS.}
\label{fig2}
\end{figure}
\end{center}

To better understand the wave group behavior consider the trajectory shown in fig. \ref{fig1} as seen in the frame of the BS. This is illustrated schematically in fig. \ref{fig2} using snapshots of the real part of the wavefunction. The spacing between wave crests of the wave group is approximately that of the center wavelength of the Gaussian distribution while snapshots for different times, shown in figs. (a)$\rightarrow$(d), illustrate motion of a wave crest.

The dashed arrows show the phase velocity for the central  Fourier component of the wave group (the other components contribute to the shape of the wave group envelope) while the solid arrows indicate the velocity of the wave group. The star symbol in this fig. follows a specific crest of the central Fourier component of the wave group as it moves through the sequence of snapshots.

For the Schr\"odinger eq. (shown in the first column of fig. \ref{fig2}) this figure is related to the lab frame in the following manner. In fig. \ref{fig2} (a) the attenuator is initially stationary in the lab frame.  In fig. \ref{fig2} (b) the wave group has passed through the attenuator while the attenuator is moving at speed $V$ and is about to overtake this transmitted wave group. In fig. \ref{fig2} (c) the attenuator has passed through the wave group in the lab frame and has come to rest while the wave group is about to traverse the attenuator again.  In fig. \ref{fig2} (d) the wave group has traversed the attenuator for the final time. The wave crest indicated by the star symbol then traverses the attenuator three times. The phase and group velocity vectors point in the same direction for all snapshots.

This, however, is not the case for the Klein-Gordon eq., illustrated in the second column of fig. \ref{fig2}. The phase velocity is large enough to ensure that its direction in the attenuator frame never reverses.

It appears in fig. \ref{fig2} that the Klein-Gordon wave crests within the wave group traverse the attenuator multiple times. However, the star symbol in the second column of fig. \ref{fig2} (a) has moved to the left away from the attenuator and out of the frames in figs. (b)$\rightarrow$(d). The wave crests that are shown in figs. (b)$\rightarrow$(d) represent later wave crests of the same Fourier component (the center wavelength of the Gaussian distribution) that move through the \ref{fig2} (a) wave group. Therefore, each wave crest of this Fourier component traverses the attenuator only once while the wave group moves through the attenuator three times.

The Schr\"odinger eq. allows the localized wave group to be interpreted as the particle's probable position, with the observed attenuation corresponding to the wave group traversing the beam splitter (BS) three times. In the Klein-Gordon eq. the wave group traverses the attenuator three times while the attenuation corresponds to only one pass through the BS. The electromagnetic wave eq. also predicts wave group attenuation for one traversal of the attenuator but in this case the attenuator cannot overtake an electromagnetic wave group traveling at the speed of light.

\subsection{Interferometer-free measurement}

Instead of using the interferometer shown in fig. \ref{fig1}, a direct measurement can be made with the same attenuator trajectory. In this case, the input-output Sagnac BS is removed. The wave group from the source then travels only in the CW direction. The detector measures the PDF of the wave group that is attenuated three times in the Schr\"odinger eq. The Klein-Gordon eq. predicts a PDF that has been attenuated by only one pass through the BS, yet the wave group passes three times through the BS.

The wave group’s shape is determined by the interference between its Fourier components. These are uniformly attenuated in the Klein-Gordon eq.. In the Schr\"odinger eq. certain components escape threefold attenuation when the attenuator moves slower than their phase velocity, thereby distorting the PDF. This demonstrates the influence that the phase velocity has on wave group dynamics. This measurement avoids interferometric methods—such as the Sagnac interferometer—making it significantly easier to implement experimentally.

\subsection{Reflected wave groups}

For the Schr\"odinger eq. the change in transmitted amplitude occurs at three distinct times; initially when the wave group traverses the static BS, then as the BS traverses the transmitted wave group, and after the BS finally comes to rest. The detector then measures a single transmitted wave group whose attenuation is cumulative.

The wave groups diffracted in first order from the grating attenuator (related to those reflected from a plate attenuator) are considered next for the Schr\"odinger equation. Let the diffracted particle flux for the stationary BS be attenuated by factor $T^{1}$ from that incident. The first diffracted wave group is generated at the initial stationary BS position. It moves away from the zeroth order transmitted wave group. As the beam splitter passes this zeroth-order transmitted wave group (attenuated by $T$), it generates a secondary first-order diffracted wave group, yielding a combined attenuation factor of $TT^{1}$. Finally, a third diffracted wave group is generated as the lagging zeroth order wave group (that has now passed twice through the attenuator) traverses the grating after it comes to rest. There are then three separate diffracted wave groups each of which has a sequentially larger attenuation.

For the Klein-Gordon eq. the first diffracted wave group is also produced at the initially stationary BS and attenuated by a factor $T^{1}$. Another diffracted wave group occurs as the BS overtakes the initial zeroth-order transmitted wave group. However, the wavecrests within this overtaken wave group did not pass through the grating while it was at rest. They originate from a later section of the wave trains within each Fourier component, forming the wave packet that is now reaching the grating. Although the BS appears to advance into the initial zeroth-order transmitted wave group, the individual crests forming this wave group actually propagate from behind the moving BS, moving into the BS (rather than the BS moving into them). Because this is their first pass through the grating, the attenuation factor of this second diffracted wave group is only $T^{1}$.

After the grating again comes to rest a third diffracted wave group is generated. The wavecrests traversing the grating at this point come from yet later segments of the incoming wave group's Fourier components—again crests that haven't yet interacted with the grating. This third diffracted wavegroup then has attenuation factor $T^{1}$. Three separate diffracted wave groups are generated along the BS trajectory, each of which with attenuation $T^{1}$.

\section{Discussion}

The boundary conditions force the amplitude change between the incident and transmitted waves to stay locked to the moving attenuator. To illustrate this consider a thin plate BS that follows the trajectory described above. No change in phase occurs for the transmitted wave. However, the reflected beam undergoes a time varying phase (a frequency shift) to satisfy the moving boundary condition.

Solutions to the Klein-Gordon eq. in the non-relativistic limit include rest mass energy whereas the Schr\"odinger eq. defines energy solely in terms of kinetic and potential components. By including rest energy, the single-particle model can account for energy transferred to internal degrees of freedom during inelastic collisions. \cite{kowalskiedu}. Modeling inelastic processes is more complex with the Schr\"odinger eq. Nevertheless, attenuation can be modeled with the Schr\"odinger eq. since only the amplitude is reduced while the particle's energy is not altered.

A constant energy $qV$ can be added to the Hamiltonian if the attenuation apparatus is completely within a hollow charged conductor. This differs from the electric Aharonov–Bohm effect, in which a charged particle enters a hollow conductor through an opening. In such a case, the particle's energy before entering is the same as that after. A constant is not added to the total energy. For an appropriate sign of V, the group velocity decreases while the phase velocity increases within the hollow conductor.

On the other hand, for the particle and attenuator apparatus that is completely within a hollow conductor (with no opening) the total energy is offset by $qV$. Depending on the sign of $V$, this term can cancel the $mc^{2}$ term in the Klein-Gordon Hamiltonian. When the potential energy cancels the rest energy, the apparatus described above will measure the same attenuation as that for the Schr\"odinger eq. However, there is no phase difference between the arms of an interferometer placed completely inside this hollow conductor since a constant added to the Hamiltonian has no effect on such a phase measurement.

The attenuation of the wavefunction, described above, illustrates an empirical gap in our knowledge of the relationship between the Schr\"odinger and non-relativistic Klein-Gordon eqs. No such gap exists in a phase measurement.

Experimental verification of the Schr\"odinger eq. would confirm that adding a constant (in this case $mc^{2}$) to the energy has no effect. It would also reveal the quantum nature of short coherence length matter without the need for interferometers. In addition, any effect of decoherence on the attenuation of a particle could be probed using an optical BS for atoms that interacts exclusively with the target body rather than surrounding gas particles.

On the other hand, experimental confirmation of Klein-Gordon behavior would highlight the breakdown of the Schr\"odinger eq., revealing relativistic effects in the non-relativistic regime. It would challenge the long-standing assumption in quantum mechanics that shifting energy by a constant has no physical effect. It would further underscore the counterintuitive nature of quantum mechanics: a particle represented as a wave group experiences a single attenuation, despite multiple passes through the attenuator.

\begin{acknowledgments}
I wish to thank Mark Lusk for thoughtful suggestions that have improved the quality of the manuscript.
\end{acknowledgments}



\end{document}